# TO THE NON-LOCAL THEORY OF THE HIGH TEMPERATURE SUPERCONDUCTIVITY

Boris V. Alexeev

Moscow Lomonosov University of Fine Chemical Technologies (MITHT)

Prospekt Vernadskogo, 86, Moscow 119570, Russia

boris.vlad.alexeev@gmail.com

The possibility of the non local physics application in the theory of superconductivity is investigated. It is shown that by the superconducting conditions the relay ("estafette") motion of the soliton' system ("lattice ion – electron") is realizing by the absence of chemical bonds. From the position of the quantum hydrodynamics the problem of creation of the high temperature superconductors leads to finding of materials which lattices could realize the soliton' motion without destruction. These materials should be created using the technology of quantum dots.

## 1. Introduction

This paper is directed on investigation of possible applications of the non-local quantum hydrodynamics in the theory of superconductivity including the high temperature superconductivity. Non-local physics demonstrates its high efficiency in many fields – from the atom structure problems to cosmology [1 - 18]. Mentioned works contain not only strict theory, but also delivering the qualitative aspects of theory without excessively cumbersome formulas. Nevertheless some remarks and explanations should be done.

As it is shown (see, for example [1, 2]) the theory of transport processes (including quantum mechanics) can be considered in the frame of unified theory based on the non-local physical description.

Let us discuss the correlation between the generalized Boltzmann physical kinetics and the laws of conservation. A rigorous description is found, for example, in the monograph [1].

Transport processes in open dissipative systems are considered in physical kinetics. Therefore, the kinetic description is inevitably related to the system diagnostics. Such an element of diagnostics in the case of theoretical description in physical kinetics is the concept of the physically infinitely small volume (**PhSV**). The correlation between theoretical description and system diagnostics is well-known in physics. Suffice it to recall the part played by test charge in electrostatics or by test circuit in the physics of magnetic phenomena. The traditional definition of **PhSV** contains the statement to the effect that the **PhSV** contains a sufficient number of particles for introducing a statistical description; however, at the same time, the **PhSV** is much smaller than the volume $V$ of the physical system under consideration; in a first approximation, this leads to local approach in investigating the transport processes. It is assumed in classical hydrodynamics that local thermodynamic equilibrium is first established within the **PhSV**, and only after that the transition occurs to global thermodynamic equilibrium if it is at all possible for the system under study.

Let us consider the hydrodynamic description in more detail from this point of view. Assume that we have two neighboring physically infinitely small volumes **PhSV₁** and **PhSV₂** in a non-equilibrium system. The one-particle distribution function (DF) $f_{sm,1}(\mathbf{r}_1, \mathbf{v}, t)$ corresponds to the volume **PhSV₁**, and the function $f_{sm,2}(\mathbf{r}_2, \mathbf{v}, t)$ — to the volume **PhSV₂**. It is assumed in a first approximation that $f_{sm,1}(\mathbf{r}_1, \mathbf{v}, t)$ does not vary within **PhSV₁**, same as $f_{sm,2}(\mathbf{r}_2, \mathbf{v}, t)$ does not vary within the neighboring volume **PhSV₂**. It is this assumption of locality that is implicitly contained in the Boltzmann equation (BE). However, the assumption is too crude. Indeed, a particle on the boundary between two volumes, which experienced the last collision in **PhSV₁** and moves toward **PhSV₂**, introduces information about the $f_{sm,1}(\mathbf{r}_1, \mathbf{v}, t)$ into the neighboring volume **PhSV₂**. Similarly, a particle on the boundary between two volumes, which experienced the last collision in **PhSV₂** and moves toward **PhSV₁**, introduces information about the DF $f_{sm,2}(\mathbf{r}_2, \mathbf{v}, t)$ into the neighboring volume **PhSV₁**. The relaxation over translational degrees of freedom of particles of like masses occurs during several collisions. As a result, "Knudsen layers" are formed on the boundary between neighboring physically infinitely small volumes, the characteristic dimension of which is of the order of path length. Therefore, a correction must be introduced into the DF in the **PhSV**, which is proportional to the mean time between collisions and to the substantive derivative of the DF being measured (rigorous derivation is given in [1]). Let a particle of finite radius be characterized as before by the position $\mathbf{r}$ at the instant of time $t$ of its center of mass moving at velocity $\mathbf{v}$. Then, the situation is possible where, at some instant of time $t$, the particle is located on the interface between two volumes. In so doing, the lead effect is possible (say, for **PhSV₂**), when the center of mass of particle moving to the neighboring volume $\mathbf{PhSV_2}$ is still in $\mathbf{PhSV_1}$. However, the delay effect takes place as well, when the center of mass of particle moving to the neighboring volume (say, $\mathbf{PhSV_2}$) is already located in $\mathbf{PhSV_2}$ but a part of the particle still belongs to $\mathbf{PhSV_1}$. This entire complex of effects defines non-local effects in space and time.

The physically infinitely small volume (**PhSV**) is an *open* thermodynamic system *for any division of macroscopic system by a set of PhSVs*. However, the BE [4]

$$Df/Dt = J^B,  \qquad (1.1)$$

where $J^B$ is the Boltzmann collision integral and $D/Dt$ is a substantive derivative, fully ignores non-local effects and contains only the local collision integral $J^B$. The foregoing nonlocal effects are insignificant only in equilibrium systems, where the kinetic approach changes to methods of statistical mechanics.

This is what the difficulties of classical Boltzmann physical kinetics arise from. Also a weak point of the classical Boltzmann kinetic theory is the treatment of the dynamic properties of interacting particles. On the one hand, as follows from the so-called "physical" derivation of the BE, Boltzmann particles are regarded as material points; on the other hand, the collision integral in the BE leads to the emergence of collision cross sections.

The rigorous approach to derivation of kinetic equation relative to one-particle DF $f$ ($KE_f$) is based on employing the hierarchy of Bogoliubov equations. Generally speaking, the structure of $KE_f$ is as follows:

$$\frac{Df}{Dt} = J^B + J^{nl},  \qquad (1.2)$$

where $J^{nl}$ is the non-local integral term. An approximation for the second collision integral is suggested by me in *generalized* Boltzmann physical kinetics,

$$J^{nl} = \frac{D}{Dt}\left(\tau \frac{Df}{Dt}\right).  \qquad (1.3)$$

Here, $\tau$ is the mean time *between* collisions of particles, which is related in a hydrodynamic approximation with dynamical viscosity $\mu$ and pressure $p$,

$$\tau\ p = \Pi\mu,  \qquad (1.4)$$

where the factor $\Pi$ is defined by the model of collision of particles; for neutral hard-sphere gas, $\Pi = 0.8$ [19]. All of the known methods of deriving kinetic equation relative to one-particle DF $f$ lead to approximation (1.3), including the method of many scales, the method of correlation functions, and the iteration method. One can draw an analogy with the Bhatnagar–Gross–Krook (BGK) approximation for local integral $J^B$,

$$J^B = \frac{f^{(0)} - f}{\tau_r}, \tag{1.5}$$

(in the simplest case the relaxation time $\tau_r \sim \tau$) the popularity of which in the case of Boltzmann collision integral is explained by the colossal simplification attained when using this approximation. The order of magnitude of the ratio between the second and first terms of the right-hand part of Eq. (1.2) is $Kn^2$, at high values of Knudsen number, these terms come to be of the same order. It would seem that, at low values of Knudsen number corresponding to hydrodynamic description, the contribution by the second term of the right-hand part of Eq. (1.2) could be ignored. However, this is not the case. Upon transition to hydrodynamic approximation (following the multiplication of the kinetic equation by invariants collision and subsequent integration with respect to velocities), the Boltzmann integral part goes to zero, and the second term of the right-hand part of Eq. (1.2) *does not go to zero* after this integration and produces a contribution of the same order in the case of generalized Navier–Stokes description.

From the mathematical standpoint, disregarding the term containing a small parameter with higher derivative is impermissible. From the physical standpoint, the arising additional terms proportional to viscosity correspond to Kolmogorov small-scale turbulence; the fluctuations are tabulated [1, 5]. It turns out that the integral term $J^{nl}$ is important from the standpoint of the theory of transport processes at both low and high values of Knudsen number. Note the treatment of GBE from the standpoint of fluctuation theory,

$$Df^a/Dt = J^B, \tag{1.6}$$

$$f^a = f - \tau Df/Dt \tag{1.7}$$

Equations (1.6) and (1.7) have a correct free-molecule limit. Therefore, $\tau Df/Dt$ is a fluctuation of distribution function, and the notation (1.6) disregarding (1.7) renders the BE open. From the standpoint of fluctuation theory, Boltzmann employed the simplest closing procedure

$$f^a = f. \tag{1.8}$$

Fluctuation effects occur in any open thermodynamic system bounded by a control surface transparent to particles. GBE (1.6) leads to generalized hydrodynamic equations [1], for example, to the continuity equation

$$\frac{\partial \rho^a}{\partial t} + \frac{\partial}{\partial \mathbf{r}} \cdot (\rho \mathbf{v}_0)^a = 0, \tag{1.9}$$

where $\rho^a$, $\mathbf{v}_0{}^a$, $(\rho \mathbf{v}_0)^a$ are calculated in view of non-locality effect in terms of gas density $\rho$, hydrodynamic velocity of flow $\mathbf{v}_0$, and density of momentum flux $\rho \mathbf{v}_0$; for locally Maxwellian distribution, $\rho^a$, $(\rho \mathbf{v}_0)^a$ are defined by the relations

$$\left(\rho - \rho^a\right)/\tau = \frac{\partial \rho}{\partial t} + \frac{\partial}{\partial \mathbf{r}} \cdot (\rho \mathbf{v}_0), \quad \left(\rho \mathbf{v}_0 - (\rho \mathbf{v}_0)^a\right)/\tau = \frac{\partial}{\partial t}(\rho \mathbf{v}_0) + \frac{\partial}{\partial \mathbf{r}} \cdot \rho \mathbf{v}_0 \mathbf{v}_0 + \bar{\bar{I}} \cdot \frac{\partial p}{\partial \mathbf{r}} - \rho \mathbf{a}, \ (1.10)$$

where $\bar{\bar{I}}$ is a unit tensor, and $\mathbf{a}$ is the acceleration due to the effect of mass forces.

In the general case, the parameter $\tau$ is the non-locality parameter; in quantum hydrodynamics, its magnitude is defined by the "time-energy" uncertainty relation [11]. The violation of Bell's inequalities [20, 21] is found for local statistical theories, and the transition to non-local description is inevitable.

The following conclusion of principal significance can be done from the previous consideration [2]:

1. Madelung's quantum hydrodynamics is equivalent to the Schrödinger equation (SE) and leads to description of the quantum particle evolution in the form of Euler equation and continuity equation. SE (and Madelung's hydrodynamics as well) are the non-dissipative theories and does not contain energy equation on principal.

2. SE is consequence of the Liouville equation as result of the local approximation of non-local equations.

3. Generalized Boltzmann physical kinetics leads to the strict approximation of non-local effects in space and time and after transmission to the local approximation leads to parameter $\tau$, which on the quantum level corresponds to the uncertainty principle "time-energy".

4. GHE lead to SE as a deep particular case of the generalized Boltzmann physical kinetics and therefore of non-local hydrodynamics. From the very beginning SE was the simplest phenomenological non-local equation corresponding to the local approximation of non-local effects.

5. Several additional facts should be noted. The local theory of turbulence based on the Navier-Stokes equations now is in blind alley. Local statistical dissipative theories are wrong on principal (J. Bell, 1964). The terms of Kn order are partly omitted in local Boltzmann equation (BE). In particular BE does not work on the distances of the radius of the particles interactions. BE belongs only to the class of plausible equations (B. Alexeev, 1982). The disaster with non-diagnostic dark energy and dark matter is obliged to local statistical description [22].

I believe that real following progress in the superconductivity (SC) can be reached only on the way of the non-local theory development. Then no reason here go into details of existing theories based on the SE apart of the facts of the principal significance leading to the non-local model of SC.

It is well known that SC was discovered more than 100 years ago in the laboratory of Heike Kamerlingh Onnes, at Leiden University in the Netherlands. It was done on 8 April 1911, after testing for electrical resistance in a sample of mercury at 3 K.

Quantum mechanics created in the 1920s provided an underlying model for the structure of ordinary metals. Metal atoms form a regular crystalline lattice with tightly bound inner core of electrons. But their loosely attached outer electrons become unbound, collecting into a mobile "electron cloud". Under the influence of an electric field, these free electrons will drift throughout the lattice, forming the basis of conductivity. But random thermal fluctuations scatter the electrons, interrupting their forward motion and dissipating energy — thereby producing electrical resistance. If some metals are cooled to temperatures close to absolute zero, the electrons suddenly shift into a highly ordered state and travel collectively without deviating from their path. Below a critical temperature $T_c$ (or better to say a narrow temperature diapason $\Delta T$) the electrical resistance falls to zero, and they become superconductors. In the case of low temperature SC $\Delta T$ is about $0,001 \div 0,1$ K, for high temperature SC $\Delta T$ can be more than 1 K. From position of non-local physics it means that the solution of the generalized non-local quantum equations transforms into soliton's type without destruction in space and time.

In the May issue of Physical Review two experimental papers [23, 24] (both received by the journal on March 24th, 1950) were describing measurements of the critical temperature of mercury for different isotopes, reporting that "there is a systematic decrease of transition temperature with increasing mass" [23], and that [24] "From these results one may infer that the transition temperature of a superconductor is a function of the nuclear mass, the lighter the mass the higher the transition temperature. From experiments follow that for crystal lattices of the different isotopes the relation takes place

$$T_c \sqrt{M} = const,\qquad\qquad(1.11)$$

where $M$ is isotopic mass and *const* is the same for all isotopes. The frequency of the lattice vibration $\omega$ is connected with the square root of $M$

$$\omega \sim 1/\sqrt{M}\ .\qquad\qquad(1.12)$$

It means that the interaction between electrons and lattice vibrations (phonons) was responsible for superconductivity.

It should be noticed that on May 16th, 1950, Herbert Fröhlich's paper entitled "Theory of the Superconducting State. I. The Ground State at the Absolute Zero of Temperature" was received by the Physical Review [25]. This theoretical paper also proposed that the interaction between electrons and lattice vibrations (phonons) was responsible for superconductivity. The paper made no mention of the experimental papers [23, 24], moreover later Herbert Fröhlich stated that the isotope effect experiments "have just come to my notice" and pointed out that the formalism in his May 16th paper ([25]) in fact predicted the effect. The question of priority is discussed until now [26].

Significant step in explanation of the behavior of superconducting materials consists in creation of BCS theory (John Bardeen, Leon N. Cooper, and John R. Schrieffer, 1957, [27]). From the qualitative point of view the physical model looks as follows. Two forces of the electric origin influence on the electrons behavior in a metal – the repulsion between electrons and attraction between electrons and positive ions that make up the rigid lattice of the metal.

Let us estimate the character scale $\Delta r$ of the phonon interaction. Phonon energy is $\hbar \omega_D \sim \hbar v_s / a$, where $\omega_D$ is Debye frequency, $v_s$ – sound speed, $a$ - the lattice constant. For example the lattice constant for a common carbon diamond is $a = 3.57 \text{Å}$ at 300 K. The character impulse is $\Delta p \sim \hbar \omega_D / v_F$, where $v_F \left( \sim 10^6 m/\sec \right)$ – electron velocity near Fermi surface. Then the scale $\Delta r$ can be found using the Heisenberg uncertainty relation

$$\Delta r \sim \hbar / \Delta p \sim v_F / \omega_D \sim \frac{v_F}{v_s} a \sim \sqrt{\frac{M}{m}} a \,, \qquad (1.13)$$

where $M$, $m$ – ion and electron masses respectively. Usually, $\Delta r \sim 10^{-5} \div 10^{-6}$ см.

The mentioned attraction distorts the ion lattice, increasing the local positive charge density of the lattice. This perturbation of the positive charge can attract other electrons. At long distances this attraction between electrons due to the displaced ions can overcome the electrons' repulsion due to their negative charge, and cause them to pair up. A Cooper pair (described in 1956 by Leon Cooper) has special construction:

1.      Electrons are fermions, but a Cooper pair is a composite boson as its total spin is integer. As result the wave functions are symmetric under particle interchange, and they are allowed to be in the same state. The Cooper pairs "condense" in a body into the same ground quantum state.

2.      If the electric current is absent, the combined impulse of a Cooper pair is equal to zero. After application of the external electric field a Cooper pair receive the additional impulse

$$\left( \mathbf{p} + \mathbf{p}' \right) + \left( -\mathbf{p} + \mathbf{p}' \right) = 2\mathbf{p}' \,, \qquad (1.14)$$

if the initial impulse of the first electron was $\mathbf{p}$ and the second one $\left( -\mathbf{p} \right)$. The impulse of the Cooper pair is

$$p_C = 2mv \qquad (1.15)$$

The electrons in a pair are not necessarily close together; because the interaction is long range. The typical estimation is $\sim 10^{-4}$ cm. This distance is usually greater than the average inter-electron distance. As result a Cooper pair begins to move as a single object under influence of the self-consistent electric field. As a superconductor is warmed, its Cooper pairs separate into individual electrons, and the material becomes no superconducting. In other words thermal energy can break the pairs. In Internet the animations can be found (see for example bcs.anim.GIF and Fig. 1.1) for illustration of the mentioned motion.

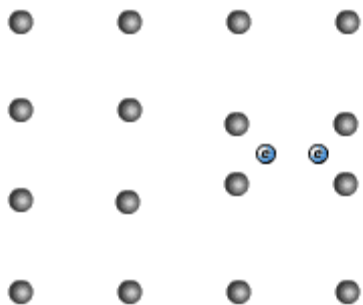

Fig. 1.1. A shot from animation illustrating a cooper pair movement (bcs.anim.GIF).

The following remarks should be taken into account by using such kind of animations:

1. The distance between electrons in a Cooper pair practically much more than the distance between neighboring ions in the crystal lattice.

2. By the concurrent motion of many Cooper pairs the lattice ions perform the oscillations along the directed motion of electrons.

From the previous consideration follows that we have quantum hydrodynamic non-local effects.

Later in June 1986, physicists Georg Bednorz and Alex Müller at the IBM Laboratory in Zurich, Switzerland, reported that they had created a material that became superconducting at 35 K. Extremely important that they were looking not at metals, but at insulating materials called copper oxides. Ceramics $YBa_2Cu_3O_7$ and TlBaCaCuO became superconducting at 93 K. For high temperature SC classic BCS theory can't be applied. Finishing the introduction is reasonable to cite the beginning of the BCS-paper [27]: "Since the discovery of the isotope effect, it has been known that superconductivity arises from the interaction between electrons and lattice vibrations, but it has proved difficult to construct an adequate theory based on this concept."

The main aim of the following consideration consists in construction "an adequate theory based on this concept", the theory distinguished from BCS.

## 2. Quantum hydrodynamic equations in the self-consistent electric field.

In general case the strict consideration leads to the following system of the non-local quantum hydrodynamic equations written in the form of the generalized Euler equations [1, 2, 10 − 13].

Continuity equation for species $\alpha$:

$$\frac{\partial}{\partial t}\left\{\rho_\alpha - \tau_\alpha\left[\frac{\partial \rho_\alpha}{\partial t} + \frac{\partial}{\partial \mathbf{r}}\cdot\left(\rho_\alpha \mathbf{v}_0\right)\right]\right\} +$$
$$+ \frac{\partial}{\partial \mathbf{r}}\cdot\left\{\rho_\alpha \mathbf{v}_0 - \tau_\alpha\left[\frac{\partial}{\partial t}\left(\rho_\alpha \mathbf{v}_0\right) + \frac{\partial}{\partial \mathbf{r}}\cdot\left(\rho_\alpha \mathbf{v}_0 \mathbf{v}_0\right) + \overset{\leftrightarrow}{\mathbf{I}}\cdot\frac{\partial p_\alpha}{\partial \mathbf{r}} - \rho_\alpha \mathbf{F}_\alpha^{(1)} - \frac{q_\alpha}{m_\alpha}\rho_\alpha \mathbf{v}_0\times\mathbf{B}\right]\right\} = R_\alpha. \tag{2.1}$$

Continuity equation for mixture:

$$\frac{\partial}{\partial t}\left\{\rho - \sum_\alpha \tau_\alpha\left[\frac{\partial \rho_\alpha}{\partial t} + \frac{\partial}{\partial \mathbf{r}}\cdot\left(\rho_\alpha \mathbf{v}_0\right)\right]\right\} + \frac{\partial}{\partial \mathbf{r}}\cdot\left\{\rho \mathbf{v}_0 - \sum_\alpha \tau_\alpha\left[\frac{\partial}{\partial t}\left(\rho_\alpha \mathbf{v}_0\right) + \frac{\partial}{\partial \mathbf{r}}\cdot\left(\rho_\alpha \mathbf{v}_0 \mathbf{v}_0\right) +\right.\right.$$
$$\left.\left. + \overset{\leftrightarrow}{\mathbf{I}}\cdot\frac{\partial p_\alpha}{\partial \mathbf{r}} - \rho_\alpha \mathbf{F}_\alpha^{(1)} - \frac{q_\alpha}{m_\alpha}\rho_\alpha \mathbf{v}_0\times\mathbf{B}\right]\right\} = 0. \tag{2.2}$$

Momentum equation for species $\alpha$:

$$\frac{\partial}{\partial t}\left\{\rho_\alpha \mathbf{v}_0 - \tau_\alpha\left[\frac{\partial}{\partial t}\left(\rho_\alpha \mathbf{v}_0\right) + \frac{\partial}{\partial \mathbf{r}}\cdot\rho_\alpha \mathbf{v}_0 \mathbf{v}_0 + \frac{\partial p_\alpha}{\partial \mathbf{r}} - \rho_\alpha \mathbf{F}_\alpha^{(1)} -\right.\right.$$
$$\left.\left. - \frac{q_\alpha}{m_\alpha}\rho_\alpha \mathbf{v}_0\times\mathbf{B}\right]\right\} - \mathbf{F}_\alpha^{(1)}\left[\rho_\alpha - \tau_\alpha\left(\frac{\partial \rho_\alpha}{\partial t} + \frac{\partial}{\partial \mathbf{r}}\left(\rho_\alpha \mathbf{v}_0\right)\right)\right] -$$
$$- \frac{q_\alpha}{m_\alpha}\left\{\rho_\alpha \mathbf{v}_0 - \tau_\alpha\left[\frac{\partial}{\partial t}\left(\rho_\alpha \mathbf{v}_0\right) + \frac{\partial}{\partial \mathbf{r}}\cdot\rho_\alpha \mathbf{v}_0 \mathbf{v}_0 + \frac{\partial p_\alpha}{\partial \mathbf{r}} - \rho_\alpha \mathbf{F}_\alpha^{(1)} -\right.\right.$$
$$\left.\left. - \frac{q_\alpha}{m_\alpha}\rho_\alpha \mathbf{v}_0\times\mathbf{B}\right]\right\}\times\mathbf{B} + \frac{\partial}{\partial \mathbf{r}}\cdot\left\{\rho_\alpha \mathbf{v}_0 \mathbf{v}_0 + p_\alpha\overset{\leftrightarrow}{\mathbf{I}} - \tau_\alpha\left[\frac{\partial}{\partial t}\left(\rho_\alpha \mathbf{v}_0 \mathbf{v}_0 +\right.\right.\right.$$
$$\left. + p_\alpha\overset{\leftrightarrow}{\mathbf{I}}\right) + \frac{\partial}{\partial \mathbf{r}}\cdot\rho_\alpha\left(\mathbf{v}_0 \mathbf{v}_0\right)\mathbf{v}_0 + 2\overset{\leftrightarrow}{\mathbf{I}}\left(\frac{\partial}{\partial \mathbf{r}}\cdot\left(p_\alpha \mathbf{v}_0\right)\right) + \frac{\partial}{\partial \mathbf{r}}\cdot\left(\overset{\leftrightarrow}{\mathbf{I}}p_\alpha \mathbf{v}_0\right) -$$
$$\left.\left. - \mathbf{F}_\alpha^{(1)}\rho_\alpha \mathbf{v}_0 - \rho_\alpha \mathbf{v}_0 \mathbf{F}_\alpha^{(1)} - \frac{q_\alpha}{m_\alpha}\rho_\alpha\left[\mathbf{v}_0\times\mathbf{B}\right]\mathbf{v}_0 - \frac{q_\alpha}{m_\alpha}\rho_\alpha \mathbf{v}_0\left[\mathbf{v}_0\times\mathbf{B}\right]\right]\right\} =$$
$$= \int m_\alpha \mathbf{v}_\alpha J_\alpha^{st,el}d\mathbf{v}_\alpha + \int m_\alpha \mathbf{v}_\alpha J_\alpha^{st,inel}d\mathbf{v}_\alpha. \tag{2.3}$$

Momentum equation for mixture

$$\frac{\partial}{\partial t}\left\{\rho\mathbf{v}_0 - \sum_\alpha \tau_\alpha\left[\frac{\partial}{\partial t}(\rho_\alpha\mathbf{v}_0) + \frac{\partial}{\partial \mathbf{r}}\cdot\rho_\alpha\mathbf{v}_0\mathbf{v}_0 + \frac{\partial p_\alpha}{\partial \mathbf{r}} - \rho_\alpha\mathbf{F}_\alpha^{(1)} - \right.\right.$$

$$\left.- \frac{q_\alpha}{m_\alpha}\rho_\alpha\mathbf{v}_0\times\mathbf{B}\right] - \sum_\alpha\mathbf{F}_\alpha^{(1)}\left[\rho_\alpha - \tau_\alpha\left(\frac{\partial\rho_\alpha}{\partial t} + \frac{\partial}{\partial \mathbf{r}}(\rho_\alpha\mathbf{v}_0)\right)\right] -$$

$$- \sum_\alpha\frac{q_\alpha}{m_\alpha}\left\{\rho_\alpha\mathbf{v}_0 - \tau_\alpha\left[\frac{\partial}{\partial t}(\rho_\alpha\mathbf{v}_0) + \frac{\partial}{\partial \mathbf{r}}\cdot\rho_\alpha\mathbf{v}_0\mathbf{v}_0 + \frac{\partial p_\alpha}{\partial \mathbf{r}} - \rho_\alpha\mathbf{F}_\alpha^{(1)} - \right.\right.$$

$$\left.\left.- \frac{q_\alpha}{m_\alpha}\rho_\alpha\mathbf{v}_0\times\mathbf{B}\right]\right\}\times\mathbf{B} + \frac{\partial}{\partial \mathbf{r}}\cdot\left\{\rho\mathbf{v}_0\mathbf{v}_0 + p\bar{\mathbf{I}} - \sum_\alpha\tau_\alpha\left[\frac{\partial}{\partial t}(\rho_\alpha\mathbf{v}_0\mathbf{v}_0 + \right.\right. \tag{2.4}$$

$$+ p_\alpha\bar{\mathbf{I}}) + \frac{\partial}{\partial \mathbf{r}}\cdot\rho_\alpha(\mathbf{v}_0\mathbf{v}_0)\mathbf{v}_0 + 2\bar{\mathbf{I}}\left(\frac{\partial}{\partial \mathbf{r}}\cdot(p_\alpha\mathbf{v}_0)\right) + \frac{\partial}{\partial \mathbf{r}}\cdot(\bar{\mathbf{I}}p_\alpha\mathbf{v}_0) -$$

$$\left.\left.- \mathbf{F}_\alpha^{(1)}\rho_\alpha\mathbf{v}_0 - \rho_\alpha\mathbf{v}_0\mathbf{F}_\alpha^{(1)} - \frac{q_\alpha}{m_\alpha}\rho_\alpha[\mathbf{v}_0\times\mathbf{B}]\mathbf{v}_0 - \frac{q_\alpha}{m_\alpha}\rho_\alpha\mathbf{v}_0[\mathbf{v}_0\times\mathbf{B}]\right]\right\} = 0$$

Energy equation for $\alpha$ species

$$\frac{\partial}{\partial t}\left\{\frac{\rho_\alpha v_0^2}{2} + \frac{3}{2}p_\alpha + \varepsilon_\alpha n_\alpha - \tau_\alpha\left[\frac{\partial}{\partial t}\left(\frac{\rho_\alpha v_0^2}{2} + \frac{3}{2}p_\alpha + \varepsilon_\alpha n_\alpha\right) + \right.\right.$$

$$\left.\left.+ \frac{\partial}{\partial \mathbf{r}}\cdot\left(\frac{1}{2}\rho_\alpha v_0^2\mathbf{v}_0 + \frac{5}{2}p_\alpha\mathbf{v}_0 + \varepsilon_\alpha n_\alpha\mathbf{v}_0\right) - \mathbf{F}_\alpha^{(1)}\cdot\rho_\alpha\mathbf{v}_0\right]\right\} +$$

$$+ \frac{\partial}{\partial \mathbf{r}}\cdot\left\{\frac{1}{2}\rho_\alpha v_0^2\mathbf{v}_0 + \frac{5}{2}p_\alpha\mathbf{v}_0 + \varepsilon_\alpha n_\alpha\mathbf{v}_0 - \tau_\alpha\left[\frac{\partial}{\partial t}\left(\frac{1}{2}\rho_\alpha v_0^2\mathbf{v}_0 + \right.\right.\right.$$

$$+ \frac{5}{2}p_\alpha\mathbf{v}_0 + \varepsilon_\alpha n_\alpha\mathbf{v}_0\Big) + \frac{\partial}{\partial \mathbf{r}}\cdot\left(\frac{1}{2}\rho_\alpha v_0^2\mathbf{v}_0\mathbf{v}_0 + \frac{7}{2}p_\alpha\mathbf{v}_0\mathbf{v}_0 + \frac{1}{2}p_\alpha v_0^2\bar{\mathbf{I}} + \right.$$

$$+ \frac{5}{2}\frac{p_\alpha^2}{\rho_\alpha}\bar{\mathbf{I}} + \varepsilon_\alpha n_\alpha\mathbf{v}_0\mathbf{v}_0 + \varepsilon_\alpha\frac{p_\alpha}{m_\alpha}\bar{\mathbf{I}}\Big) - \rho_\alpha\mathbf{F}_\alpha^{(1)}\cdot\mathbf{v}_0\mathbf{v}_0 - p_\alpha\mathbf{F}_\alpha^{(1)}\cdot\bar{\mathbf{I}} -$$

$$- \frac{1}{2}\rho_\alpha v_0^2\mathbf{F}_\alpha^{(1)} - \frac{3}{2}\mathbf{F}_\alpha^{(1)}p_\alpha - \frac{\rho_\alpha v_0^2}{2}\frac{q_\alpha}{m_\alpha}[\mathbf{v}_0\times\mathbf{B}] - \frac{5}{2}p_\alpha\frac{q_\alpha}{m_\alpha}[\mathbf{v}_0\times\mathbf{B}] -$$

$$- \varepsilon_\alpha n_\alpha\frac{q_\alpha}{m_\alpha}[\mathbf{v}_0\times\mathbf{B}] - \varepsilon_\alpha n_\alpha\mathbf{F}_\alpha^{(1)}\Big]\Big\} - \Big\{\rho_\alpha\mathbf{F}_\alpha^{(1)}\cdot\mathbf{v}_0 - \tau_\alpha\Big[\mathbf{F}_\alpha^{(1)}\cdot$$

$$\cdot\left(\frac{\partial}{\partial t}(\rho_\alpha\mathbf{v}_0) + \frac{\partial}{\partial \mathbf{r}}\cdot\rho_\alpha\mathbf{v}_0\mathbf{v}_0 + \frac{\partial}{\partial \mathbf{r}}\cdot p_\alpha\bar{\mathbf{I}} - \rho_\alpha\mathbf{F}_\alpha^{(1)} - q_\alpha n_\alpha[\mathbf{v}_0\times\mathbf{B}]\right)\Big]\Big\}\Big\} = \tag{2.5}$$

$$= \int\left(\frac{m_\alpha v_\alpha^2}{2} + \varepsilon_\alpha\right)J_\alpha^{st,el}d\mathbf{v}_\alpha + \int\left(\frac{m_\alpha v_\alpha^2}{2} + \varepsilon_\alpha\right)J_\alpha^{st,inel}d\mathbf{v}_\alpha.$$

Energy equation for mixture:

$$\frac{\partial}{\partial t}\left\{\frac{\rho v_0^2}{2}+\frac{3}{2}p+\sum_\alpha \varepsilon_\alpha n_\alpha -\sum_\alpha \tau_\alpha\left[\frac{\partial}{\partial t}\left(\frac{\rho_\alpha v_0^2}{2}+\frac{3}{2}p_\alpha+\varepsilon_\alpha n_\alpha\right)+\right.\right.$$

$$+\frac{\partial}{\partial \mathbf{r}}\cdot\left(\frac{1}{2}\rho_\alpha v_0^2\mathbf{v}_0+\frac{5}{2}p_\alpha\mathbf{v}_0+\varepsilon_\alpha n_\alpha\mathbf{v}_0\right)-\mathbf{F}_\alpha^{(1)}\cdot\rho_\alpha\mathbf{v}_0\left.\right]\right\}+$$

$$+\frac{\partial}{\partial \mathbf{r}}\cdot\left\{\frac{1}{2}\rho v_0^2\mathbf{v}_0+\frac{5}{2}p\mathbf{v}_0+\mathbf{v}_0\sum_\alpha\varepsilon_\alpha n_\alpha-\sum_\alpha\tau_\alpha\left[\frac{\partial}{\partial t}\left(\frac{1}{2}\rho_\alpha v_0^2\mathbf{v}_0+\right.\right.\right.$$

$$+\frac{5}{2}p_\alpha\mathbf{v}_0+\varepsilon_\alpha n_\alpha\mathbf{v}_0\right)+\frac{\partial}{\partial \mathbf{r}}\cdot\left(\frac{1}{2}\rho_\alpha v_0^2\mathbf{v}_0\mathbf{v}_0+\frac{7}{2}p_\alpha\mathbf{v}_0\mathbf{v}_0+\frac{1}{2}p_\alpha v_0^2\vec{\mathbf{I}}+\right.$$

$$+\frac{5}{2}\frac{p_\alpha^2}{\rho_\alpha}\vec{\mathbf{I}}+\varepsilon_\alpha n_\alpha\mathbf{v}_0\mathbf{v}_0+\varepsilon_\alpha\frac{p_\alpha}{m_\alpha}\vec{\mathbf{I}}\right)-\rho_\alpha\mathbf{F}_\alpha^{(1)}\cdot\mathbf{v}_0\mathbf{v}_0-p_\alpha\mathbf{F}_\alpha^{(1)}\cdot\vec{\mathbf{I}}-$$

$$-\frac{1}{2}\rho_\alpha v_0^2\mathbf{F}_\alpha^{(1)}-\frac{3}{2}\mathbf{F}_\alpha^{(1)}p_\alpha-\frac{\rho_\alpha v_0^2}{2}\frac{q_\alpha}{m_\alpha}\left[\mathbf{v}_0\times\mathbf{B}\right]-\frac{5}{2}p_\alpha\frac{q_\alpha}{m_\alpha}\left[\mathbf{v}_0\times\mathbf{B}\right]-$$

$$-\varepsilon_\alpha n_\alpha\frac{q_\alpha}{m_\alpha}\left[\mathbf{v}_0\times\mathbf{B}\right]-\varepsilon_\alpha n_\alpha\mathbf{F}_\alpha^{(1)}\left.\left.\right]\right\}-\left\{\mathbf{v}_0\cdot\sum_\alpha\rho_\alpha\mathbf{F}_\alpha^{(1)}-\right.$$

$$-\sum_\alpha\tau_\alpha\left[\mathbf{F}_\alpha^{(1)}\cdot\left(\frac{\partial}{\partial t}\left(\rho_\alpha\mathbf{v}_0\right)+\frac{\partial}{\partial \mathbf{r}}\cdot\rho_\alpha\mathbf{v}_0\mathbf{v}_0+\frac{\partial}{\partial \mathbf{r}}\cdot p_\alpha\vec{\mathbf{I}}-\rho_\alpha\mathbf{F}_\alpha^{(1)}-q_\alpha n_\alpha\left[\mathbf{v}_0\times\mathbf{B}\right]\right)\right]\right\}=0.$$

(2.6)

Here $\mathbf{F}_\alpha^{(1)}$ are the forces of the non-magnetic origin, $\mathbf{B}$ - magnetic induction, $\vec{\mathbf{I}}$ - unit tensor, $q_\alpha$ - charge of the $\alpha$-component particle, $p_\alpha$ - static pressure for $\alpha$-component, $\varepsilon_\alpha$ - internal energy for the particles of $\alpha$-component, $\mathbf{v}_0$ - hydrodynamic velocity for mixture, $\tau_\alpha$ - non-local parameter.

On principal GHE (and therefore GQH) needn't in using of the "time-energy" uncertainty relation for estimation of the value of the non-locality parameter $\tau$. Moreover the "time-energy" uncertainty relation does not produce the exact relations and from position of non-local physics is only the simplest estimation of the non-local effects. Really, let us consider two neighboring physically infinitely small volumes **PhSV₁** and **PhSV₂** in a non-equilibrium system. Obviously the time $\tau$ should tends to diminish with increasing of the velocities $u$ of particles invading in the nearest neighboring physically infinitely small volume ( **PhSV₁** or **PhSV₂** ):

$$\tau = H / u^n .$$

(2.7)

But the value $\tau$ cannot depend on the velocity direction and naturally to tie $\tau$ with the particle kinetic energy, then

$$\tau = H \big/ mu^2 ,$$

(2.8)

where $H$ is a coefficient of proportionality, which reflects the state of physical system. In the simplest case $H$ is equal to Plank constant $\hbar$ and relation (2.8) became compatible with the Heisenberg relation.

In the following we intend to obtain the soliton's type of solution of the generalized hydrodynamic equations (GHE). Much more examples of the theory applications can be found in [1 - 3]. The non-stationary 1D model will be used with taking into account the energy equation, external forces and non-locality parameter $\tau$ defined by the "time-energy" uncertainty relation of Heisenberg. Then GHE contain Poisson equation (reflected fluctuations of charges and flux of the charges density), two continuity equations for positive (lattice ions) and negative (electrons) species, momentum equation and two energy equations for positive and negative species. This system of six non-stationary 1D equations is written as:

(Generalized Poisson equation):

$$\frac{\partial^2 \varphi}{\partial x^2} = -4\pi e \left\{ \left[ n_i - \tau_i \left( \frac{\partial n_i}{\partial t} + \frac{\partial}{\partial x}(n_i u) \right) \right] - \left[ n_e - \tau_e \left( \frac{\partial n_e}{\partial t} + \frac{\partial}{\partial x}(n_e u) \right) \right] \right\}. \tag{2.9}$$

(Continuity equation for ions):

$$\frac{\partial}{\partial t} \left\{ \rho_i - \tau_i \left[ \frac{\partial \rho_i}{\partial t} + \frac{\partial}{\partial x}(\rho_i u) \right] \right\} + \frac{\partial}{\partial x} \left\{ \rho_i u - \tau_i \left[ \frac{\partial}{\partial t}(\rho_i u) + \frac{\partial}{\partial x}(\rho_i u^2) + \frac{\partial p_i}{\partial x} - \rho_i F_i \right] \right\} = 0. \tag{2.10}$$

(Continuity equation for electrons):

$$\frac{\partial}{\partial t} \left\{ \rho_e - \tau_e \left[ \frac{\partial \rho_e}{\partial t} + \frac{\partial}{\partial x}(\rho_e u) \right] \right\} + \frac{\partial}{\partial x} \left\{ \rho_e u - \tau_e \left[ \frac{\partial}{\partial t}(\rho_e u) + \frac{\partial}{\partial x}(\rho_e u^2) + \frac{\partial p_e}{\partial x} - \rho_e F_e \right] \right\} = 0. \tag{2.11}$$

(Momentum equation):

$$\frac{\partial}{\partial t} \left\{ \rho u - \tau_i \left[ \frac{\partial}{\partial t}(\rho_i u) + \frac{\partial}{\partial x}(p_i + \rho_i u^2) - \rho_i F_i \right] - \tau_e \left[ \frac{\partial}{\partial t}(\rho_e u) + \frac{\partial}{\partial x}(p_e + \rho_e u^2) - \rho_e F_e \right] \right\} -$$

$$- \rho_i F_i - \rho_e F_e + F_i \tau_i \left( \frac{\partial \rho_i}{\partial t} + \frac{\partial}{\partial x}(\rho_i u) \right) + F_e \tau_e \left( \frac{\partial \rho_e}{\partial t} + \frac{\partial}{\partial x}(\rho_e u) \right) +$$

$$+ \frac{\partial}{\partial x} \left\{ \begin{array}{l} \rho u^2 + p - \tau_i \left[ \frac{\partial}{\partial t}(\rho_i u^2 + p_i) + \frac{\partial}{\partial x}(\rho_i u^3 + 3 p_i u) - 2\rho_i u F_i \right] - \\ - \tau_e \left[ \frac{\partial}{\partial t}(\rho_e u^2 + p_e) + \frac{\partial}{\partial x}(\rho_e u^3 + 3 p_e u) \right] - 2\rho_e u F_e \end{array} \right\} = 0. \tag{2.12}$$

(Energy equation for ions):

$$\frac{\partial}{\partial t} \left\{ \rho_i u^2 + 3 p_i - \tau_i \left[ \frac{\partial}{\partial t}(\rho_i u^2 + 3 p_i) + \frac{\partial}{\partial x}(\rho_i u^3 + 5 p_i u) - 2\rho_i F_i u \right] \right\} +$$

$$+ \frac{\partial}{\partial x} \left\{ \rho_i u^3 + 5 p_i u - \tau_i \left[ \begin{array}{l} \frac{\partial}{\partial t}(\rho_i u^3 + 5 p_i u) + \frac{\partial}{\partial x}\left( \rho_i u^4 + 8 p_i u^2 + 5 \frac{p_i^2}{\rho_i} \right) - \\ - F_i \left( 3\rho_i u^2 + 5 p_i \right) \end{array} \right] \right\} -$$

$$- 2u\rho_i F_i + 2\tau_i F_i \left[ \frac{\partial}{\partial t}(\rho_i u) + \frac{\partial}{\partial x}(\rho_i u^2 + p_i) - \rho_i F_i \right] = -\frac{p_i - p_e}{\tau_{ei}}. \tag{2.13}$$

(Energy equation for electrons):

$$\frac{\partial}{\partial t}\left\{\rho_e u^2 + 3p_e - \tau_e\left[\frac{\partial}{\partial t}\left(\rho_e u^2 + 3p_e\right) + \frac{\partial}{\partial x}\left(\rho_e u^3 + 5p_e u\right) - 2\rho_e F_e u\right]\right\} +$$

$$+ \frac{\partial}{\partial x}\left\{\rho_e u^3 + 5p_e u - \tau_e\left[\begin{array}{l}\frac{\partial}{\partial t}\left(\rho_e u^3 + 5p_e u\right) + \frac{\partial}{\partial x}\left(\rho_e u^4 + 8p_e u^2 + 5\frac{p_e^2}{\rho_e}\right) - \\ - F_e\left(3\rho_e u^2 + 5p_e\right)\end{array}\right]\right\} - \qquad (2.14)$$

$$- 2u\rho_e F_e + 2\tau_e F_e\left[\frac{\partial}{\partial t}\left(\rho_e u\right) + \frac{\partial}{\partial x}\left(\rho_e u^2 + p_e\right) - \rho_e F_e\right] = -\frac{p_e - p_i}{\tau_{ei}},$$

where $u$ is velocity of the directed motion of combined quantum object (phonon-electron), $n_i$ and $n_e$ – numerical density of the charged species, $F_i$ and $F_e$ – forces (of potential and non-potential origin), acting on the mass unit of the charged particles. The right hand sides of the energy equations are written in the relaxation forms following from BGK kinetic approximation (compare with (1.5)).

For acting potential forces of the electrical origin the relations are valid

$$F_i^{(pot)} = -\frac{e}{m_i}\frac{\partial\varphi}{\partial x}, \ \ F_e^{(pot)} = \frac{e}{m_e}\frac{\partial\varphi}{\partial x}, \qquad (2.15)$$

where $\varphi$ – scalar potential.

Introduce approximations for $\tau_i$ and $\tau_i$ using (2.8)

$$\tau_i = \hbar\big/m_i u^2, \ \ \tau_e = \hbar\big/m_e u^2. \qquad (2.16)$$

For electron-phonon non-local parameter $\tau_{ei}$ the following relation is applicable

$$\frac{1}{\tau_{ei}} = \frac{1}{\tau_e} + \frac{1}{\tau_i}. \qquad (2.17)$$

For this case parameter $\tau_{ei}$ corresponds to the relaxation time for the positive and negative species and to Heisenberg relation for combined particle. Really

$$\frac{1}{\tau_{ei}} = \frac{\tau_e + \tau_i}{\tau_e \tau_i} = \frac{\dfrac{\hbar}{m_e u^2} + \dfrac{\hbar}{m_i u^2}}{\dfrac{\hbar^2}{u^4}\dfrac{1}{m_e m_i}} = \frac{u^2}{\hbar}\left(m_e + m_i\right). \qquad (2.18)$$

Then

$$u^2\left(m_e + m_i\right)\tau_{ei} = \hbar. \qquad (2.19)$$

Formula (2.19) is obvious consequence of uncertainty relation for combined particle which mass is $m_i + m_e$.

Energy equation of the generalized quantum hydrodynamics contain pressures $p_i, p_e$, which can be named as the quantum pressure of the non-local origin. In the definite sense these pressures can be considered as analog of the Bose condensate pressure.

### 3. Combined quantum solitons in the self-consistent electric field.

Let us formulate the problem in detail. The non-stationary 1D motion of the combined phonon-electron soliton is considered under influence of the self-consistent electric forces of the potential and non-potential origin. It should be shown that mentioned soliton can exists without a chemical bond formation. For better understanding of situation let us investigate the situation for the case when the external forces are absent. Introduce the coordinate system moving along the positive direction of the $x$ axis in 1D space with the velocity $C = u_0$, which is equal to the phase velocity of this quantum object.

$$\xi = x - Ct . \qquad (3.1)$$

Taking into account de Broglie relation we should wait that the group velocity $u_g$ is equal to $2u_0$.

Really the energy of a relativistic particle is

$$E = mc^2 , \qquad (3.2)$$

where

$$m = m_0 \left(1 - \frac{v_g^2}{c^2}\right)^{-1/2} , \qquad (3.3)$$

and $c$ is the light velocity, $v_g$ is the group velocity, $m_0$ – the particle rest mass. Relation (3.2) can be written as

$$E = p \frac{c^2}{v_g} , \qquad (3.4)$$

where

$$p = m v_g \qquad (3.5)$$

is the particle impulse. In the non-relativistic approach the relation (3.4) takes the form

$$E = \frac{1}{2} m_0 v_g^2 . \qquad (3.6)$$

Using the dualism principle in the de Broglie interpretation we have for the particle energy

$$E = \hbar\omega = \hbar k v_{ph}, \tag{3.7}$$

where $\omega$ is the circular frequency, $v_{ph} = \dfrac{\omega}{\kappa}$ – the phase velocity, $\kappa = 2\pi / \lambda$ is the wave number and $\lambda$ is the wave length. Correspondingly the particle impulse $p$ is

$$p = \hbar k \tag{3.8}$$

and using (3.7),

$$E = p v_{ph}. \tag{3.9}$$

Then in the non-relativistic case

$$E = \frac{1}{2} m_0 v_g^2 = \frac{1}{2} p v_g. \tag{3.10}$$

From (3.9) and (3.10) for the non-relativistic case one obtains (compare with (1.15))

$$v_g = 2 v_{ph}. \tag{3.11}$$

Then we should wait that the indestructible soliton has the velocity $v_{ph}$ in the coordinate system moving with the phase velocity $v_{ph}$.

If we pass on the moving coordinate system, all dependent hydrodynamic values will be functions of $(\xi, t)$. But we investigate the possibility of the creation of the combined quantum object of the soliton type. For this case the explicit time dependence of solutions does not exist in mentioned coordinate system moving with the phase velocity $u_0$.

Write down the system of equations (2.9) - (2.14) for the two component mixture of charged particles without taking into account the component's internal energy in the dimensionless form, where dimensionless symbols are marked by tildes. We begin with introduction the scales for velocity

$$[u] = u_0 \tag{3.12}$$

and for coordinate $x$

$$\frac{\hbar}{m_e u_0} = x_0. \tag{3.13}$$

Generalized Poisson equation (2.9)

$$\frac{\partial^2 \varphi}{\partial x^2} = -4\pi e \left\{ \left[ n_i - \frac{\hbar}{m_i u^2} u_0 \left( -\frac{\partial n_i}{\partial x} + \frac{\partial}{\partial x} (n_i \tilde{u}) \right) \right] - \left[ n_e - \frac{\hbar}{m_e u^2} u_0 \left( -\frac{\partial n_e}{\partial x} + \frac{\partial}{\partial x} (n_e \tilde{u}) \right) \right] \right\} \tag{3.14}$$

takes the form

$$\frac{\partial^2 \tilde{\varphi}}{\partial \tilde{\xi}^2} = -\left\{ \frac{m_e}{m_i} \left[ \tilde{\rho}_i - \frac{1}{\tilde{u}^2} \frac{m_e}{m_i} \left( -\frac{\partial \tilde{\rho}_i}{\partial \tilde{\xi}} + \frac{\partial}{\partial \tilde{\xi}} (\tilde{\rho}_i \tilde{u}) \right) \right] - \left[ \tilde{\rho}_e - \frac{1}{\tilde{u}^2} \left( -\frac{\partial \tilde{\rho}_e}{\partial \tilde{\xi}} + \frac{\partial}{\partial \tilde{\xi}} (\tilde{\rho}_e \tilde{u}) \right) \right] \right\}, \tag{3.15}$$

if the potential scale $\varphi_0$ and the density scale $\rho_0$ are chosen as ( $e$ is absolute electron charge)

$$\varphi_0 = \frac{m_e}{e} u_0^2, \qquad (3.16)$$

$$\rho_0 = \frac{m_e^4}{4\pi\hbar^2 e^2} u_0^4. \qquad (3.17)$$

Scaling forces are

$$\rho_i F_i = -\frac{u_0^2}{x_0} \rho_0 \frac{m_e}{m_i} \frac{\partial \tilde{\varphi}}{\partial \tilde{\xi}} \tilde{\rho}_i, \qquad (3.18)$$

$$\rho_e F_e = \frac{u_0^2}{x_0} \rho_0 \frac{\partial \tilde{\varphi}}{\partial \tilde{\xi}} \tilde{\rho}_e. \qquad (3.19)$$

Strictly speaking we should take into account the self-consistent magnetic field. By the Lorentz normalization the classic Poisson equation is

$$\Delta \varphi - \frac{\varepsilon \mu}{c^2} \frac{\partial^2 \varphi}{\partial t^2} = -\frac{4\pi}{\varepsilon} \rho, \qquad (3.20)$$

where $\varepsilon$, $\mu$ are dielectric and magnetic permeability and $c$ – the light velocity. As we see, in Eq. (3.20) the second time derivative is omitted as small value of the $u_0^2/c^2$ order. In relations (3.18), (3.19) the derivatives of the vector potential $\mathbf{A}$ on time are also omitted as the small values of the $u_0/c$ order. Then here the self consistent and external magnetic fields are not considered.

Analogical transformations should be applied to other equations of the system (2.9) – (2.14). As result one obtains the six non-linear dimensionless ordinary differential equations

$$\frac{\partial^2 \tilde{\varphi}}{\partial \tilde{\xi}^2} = -\left\{ \frac{m_e}{m_i} \left[ \tilde{\rho}_i - \frac{1}{\tilde{u}^2} \frac{m_e}{m_i} \left( -\frac{\partial \tilde{\rho}_i}{\partial \tilde{\xi}} + \frac{\partial}{\partial \tilde{\xi}} (\tilde{\rho}_i \tilde{u}) \right) \right] - \left[ \tilde{\rho}_e - \frac{1}{\tilde{u}^2} \left( -\frac{\partial \tilde{\rho}_e}{\partial \tilde{\xi}} + \frac{\partial}{\partial \tilde{\xi}} (\tilde{\rho}_e \tilde{u}) \right) \right] \right\}, \qquad (3.21)$$

$$\frac{\partial \tilde{\rho}_i}{\partial \tilde{\xi}} - \frac{\partial \tilde{\rho}_i \tilde{u}}{\partial \tilde{\xi}} + \frac{m_e}{m_i} \frac{\partial}{\partial \tilde{\xi}} \left\{ \frac{1}{\tilde{u}^2} \left[ \frac{\partial}{\partial \tilde{\xi}} \left( \tilde{p}_i + \tilde{\rho}_i + \tilde{\rho}_i \tilde{u}^2 - 2\tilde{\rho}_i \tilde{u}_i \right) + \frac{m_e}{m_i} \tilde{\rho}_i \frac{\partial \tilde{\varphi}}{\partial \tilde{\xi}} \right] \right\} = 0, \qquad (3.22)$$

$$\frac{\partial \tilde{\rho}_e}{\partial \tilde{\xi}} - \frac{\partial \tilde{\rho}_e \tilde{u}}{\partial \tilde{\xi}} + \frac{\partial}{\partial \tilde{\xi}} \left\{ \frac{1}{\tilde{u}^2} \left[ \frac{\partial}{\partial \tilde{\xi}} \left( \tilde{p}_e + \tilde{\rho}_e + \tilde{\rho}_e \tilde{u}^2 - 2\tilde{\rho}_e \tilde{u}_e \right) - \tilde{\rho}_e \frac{\partial \tilde{\varphi}}{\partial \tilde{\xi}} \right] \right\} = 0, \qquad (3.23)$$

$$\frac{\partial}{\partial \tilde{\xi}}\left\{(\tilde{\rho}_i + \tilde{\rho}_e)\tilde{u}^2 + (\tilde{p}_i + \tilde{p}_e) - (\tilde{\rho}_i + \tilde{\rho}_e)\tilde{u}\right\} +$$

$$+ \frac{\partial}{\partial \tilde{\xi}}\left\{
\begin{array}{l}
\dfrac{1}{\tilde{u}^2}\dfrac{m_e}{m_i}\left[\dfrac{\partial}{\partial \tilde{\xi}}\left(2\,\tilde{p}_i + 2\,\tilde{\rho}_i\tilde{u}^2 - \tilde{\rho}_i\tilde{u} - \tilde{\rho}_i\tilde{u}^3 - 3\,\tilde{p}_i\tilde{u}\right) + \tilde{\rho}_i\,\dfrac{m_e}{m_i}\dfrac{\partial \tilde{\varphi}}{\partial \tilde{\xi}}\right] + \\[2mm]
+ \dfrac{1}{\tilde{u}^2}\left[\dfrac{\partial}{\partial \tilde{\xi}}\left(2\,\tilde{p}_e + 2\,\tilde{\rho}_e\tilde{u}^2 - \tilde{\rho}_e\tilde{u} - \tilde{\rho}_e\tilde{u}^3 - 3\,\tilde{p}_e\tilde{u}\right) - \tilde{\rho}_e\,\dfrac{\partial \tilde{\varphi}}{\partial \tilde{\xi}}\right]
\end{array}
\right\} +$$

$$+ \tilde{\rho}_i\,\frac{m_e}{m_i}\frac{\partial \tilde{\varphi}}{\partial \tilde{\xi}} - \tilde{\rho}_e\,\frac{\partial \tilde{\varphi}}{\partial \tilde{\xi}} - \frac{\partial \tilde{\varphi}}{\partial \tilde{\xi}}\frac{1}{\tilde{u}^2}\left(\frac{m_e}{m_i}\right)^2\left(-\frac{\partial \tilde{\rho}_i}{\partial \tilde{\xi}} + \frac{\partial}{\partial \tilde{\xi}}(\tilde{\rho}_i\tilde{u})\right) +$$

$$+ \frac{\partial \tilde{\varphi}}{\partial \tilde{\xi}}\frac{1}{\tilde{u}^2}\left(-\frac{\partial \tilde{\rho}_e}{\partial \tilde{\xi}} + \frac{\partial}{\partial \tilde{\xi}}(\tilde{\rho}_e\tilde{u})\right) - 2\frac{\partial}{\partial \tilde{\xi}}\left\{\frac{1}{\tilde{u}}\frac{\partial \tilde{\varphi}}{\partial \tilde{\xi}}\left[\left(\frac{m_e}{m_i}\right)^2\tilde{\rho}_i - \tilde{\rho}_e\right]\right\} = 0,$$

(3.24)

$$\frac{\partial}{\partial \tilde{\xi}}\left\{\tilde{\rho}_i\tilde{u}^3 + 5\tilde{p}_i\tilde{u} - \tilde{\rho}_i\tilde{u}^2 - 3\tilde{p}_i\right\} + \frac{\partial}{\partial \xi}\left\{\frac{1}{\tilde{u}^2}\frac{m_e}{m_i}\left[\frac{\partial}{\partial \tilde{\xi}}\left(2\tilde{\rho}_i\tilde{u}^3 + 10\tilde{p}_i\tilde{u} - \tilde{\rho}_i\tilde{u}^4 - 8\tilde{p}_i\tilde{u}^2 - \right.\right.\right.$$

$$\left.\left.\left. - 5\frac{\tilde{p}_i^{\,2}}{\tilde{\rho}_i} - \tilde{\rho}_i\tilde{u}^2 - 3\tilde{p}_i\right) + \frac{m_e}{m_i}\frac{\partial \tilde{\varphi}}{\partial \tilde{\xi}}(2\tilde{\rho}_i\tilde{u} - 3\tilde{\rho}_i\tilde{u}^2 - 5\tilde{p}_i)\right]\right\} + 2\frac{m_e}{m_i}\tilde{\rho}_i\tilde{u}\frac{\partial \tilde{\varphi}}{\partial \tilde{\xi}} -$$

$$- 2\frac{\partial \tilde{\varphi}}{\partial \xi}\frac{1}{\tilde{u}^2}\left(\frac{m_e}{m_i}\right)^2\left[\frac{\partial}{\partial \tilde{\xi}}(\tilde{\rho}_i\tilde{u}^2 + \tilde{p}_i - \tilde{\rho}_i\tilde{u}) + \tilde{\rho}_i\frac{m_e}{m_i}\frac{\partial \tilde{\varphi}}{\partial \tilde{\xi}}\right] =$$

$$= -(\tilde{p}_i - \tilde{p}_e)\tilde{u}^2\left(1 + \frac{m_i}{m_e}\right)$$

(3.25)

$$\frac{\partial}{\partial \tilde{\xi}}\left\{\tilde{\rho}_e\tilde{u}^3 + 5\tilde{p}_e\tilde{u} - \tilde{\rho}_e\tilde{u}^2 - 3\tilde{p}_e\right\} + \frac{\partial}{\partial \tilde{\xi}}\left\{\frac{1}{\tilde{u}^2}\left[\frac{\partial}{\partial \tilde{\xi}}\left(2\tilde{\rho}_e\tilde{u}^3 + 10\tilde{p}_e\tilde{u} - \tilde{\rho}_e\tilde{u}^4 - 8\tilde{p}_e\tilde{u}^2 - \right.\right.\right.$$

$$\left.\left.\left. - 5\frac{\tilde{p}_e^{\,2}}{\tilde{\rho}_e} - \tilde{\rho}_e\tilde{u}^2 - 3\tilde{p}_e\right) + \frac{\partial \tilde{\varphi}}{\partial \tilde{\xi}}(3\tilde{\rho}_e\tilde{u}^2 + 5\tilde{p}_e - 2\tilde{\rho}_e\tilde{u})\right]\right\} - 2\tilde{\rho}_e\tilde{u}\frac{\partial \tilde{\varphi}}{\partial \tilde{\xi}} +$$

$$+ 2\frac{\partial \tilde{\varphi}}{\partial \tilde{\xi}}\frac{1}{\tilde{u}^2}\left[\frac{\partial}{\partial \tilde{\xi}}(\tilde{\rho}_e\tilde{u}^2 + \tilde{p}_e - \tilde{\rho}_e\tilde{u}) - \tilde{\rho}_e\frac{\partial \tilde{\varphi}}{\partial \tilde{\xi}}\right] = -(\tilde{p}_e - \tilde{p}_i)\left(1 + \frac{m_i}{m_e}\right)\tilde{u}^2$$

(3.26)

Some comments to Eqs. (3.21 – 3.26):

1. Every equation from the system (3.21 – 3.26) is of the second order and needs two conditions. The problem belongs to the class of Cauchy problems.

2. In comparison with the Schrödinger theory connected with behavior of the wave function, no special conditions are applied for dependent variables including the domain of the solution existing. This domain is defined automatically in the process of the numerical solution of the concrete variant of calculations.

3. From the introduced scales

$$u_0, \ x_0 = \frac{\hbar}{m_e}\frac{1}{u_0}, \ \varphi_0 = \frac{m_e}{e}u_0^2, \ \rho_0 = \frac{m_e^4}{4\pi\hbar^2 e^2}u_0^4, \ p_0 = \rho_0 u_0^2 = \frac{m_e^4}{4\pi\hbar^2 e^2}u_0^6$$

only one parameter is independent – the phase velocity $u_0$ of the combined quantum object. From this point of view the obtained solutions which will be discussed below have the universal character defined only by Cauchy conditions.

4. Introduced scales have the connection with the character values in the Schrödinger theory of the hydrogen atom. Really in the Schrödinger theory the probability maximum corresponds to the Bohr's orbits and for the state $1s$ the orbit radius $a = \frac{\hbar^2}{me^2} \cong 0.53\cdot10^{-8}$ cm, and the orbit velocity $V_{\text{orb}} = \frac{e}{\sqrt{m_e a}} = \frac{e^2}{\hbar} = \frac{\hbar}{m_e a}$. For example taking as the velocity scale $u_0 = V_{\text{orb}}$ we have for the numerical density scale $n_0 = \frac{\rho_0}{m_e} = \frac{m_e^3}{4\pi\hbar^2 e^2}u_0^4 = \frac{\hbar^2}{4\pi e^2 m_e a^4} = \frac{1}{4\pi a^3}$. But the probability density for the Schrödinger atom at the origin of the coordinates for 1s state corresponds to value $\left(\Psi_{100}\right)^2_{r=0} = \frac{1}{\pi a^3}$. Then for this scale choice (with an accuracy of numerical coefficient) $n_0 \approx \left(\Psi_{100}\right)^2_{r=0}$. The difference is connected with the choice of the object geometry and coordinate system.

## 4. The results of the mathematical modeling.

The system of generalized quantum hydrodynamic equations (3.21) – (3.26) have the great possibilities of mathematical modeling as result of changing of twelve Cauchy conditions describing the character features of initial perturbations which lead to the soliton formation.

On this step of investigation I intend to demonstrate the influence of difference conditions on the soliton formation. The following figures reflect some results of calculations realized according to the system of equations (3.21) - (3.26) with the help of Maple 9 (and the older Maple versions). The following notations on figures are used: r- density $\tilde{\rho}_i$, s- density $\tilde{\rho}_e$ (solid lines), u- velocity $\tilde{u}$ (dashed line), p - pressure $\tilde{p}_i$, q- pressure $\tilde{p}_e$ and v - self consistent potential $\tilde{\varphi}$. Explanations placed under all following figures, Maple program contains Maple's notations – for example the expression $D(u)(0) = 0$ means in usual notations $\frac{\partial \tilde{u}}{\partial \tilde{\xi}}(0) = 0$, independent variable $t$ responds to $\tilde{\xi}$.

There is the problem of principle significance – is it possible after a perturbation (defined by Cauchy conditions) to obtain the quantum object of the soliton's kind as result of the self-organization of ionized matter? In the case of the positive answer, what is the origin of existence of this stable object? These questions are considered in detail in [1, 2, 10-13], for Cartesian coordinate system the self-consistent calculations in the spherical coordinate system can be found in [22]. The conclusion was formulated:

1.  The stability of the quantum object is result of the self-consistent influence of electric forces and quantum pressures.

2.  In the absence of the external electric field the combined soliton corresponding to atom structure can exists if $\tilde{p}_i(0) \neq \tilde{p}_e(0)$. This inequality corresponds to energy of chemical bond.

3.  Stability can be achieved if soliton has *negative* shell and *positive* nuclei and $\tilde{p}_i(0) > \tilde{p}_e(0)$. Increasing the difference $p_i(0) - p_e(0)$ lead to diminishing of the character domain occupied by soliton. Stability can be also achieved if soliton has *positive* shell and *negative* kernel but $\tilde{p}_i(0) < \tilde{p}_e(0)$.

For illustration of item 2 let us consider the calculations when the energy of the chemical bond is equal to zero, $\tilde{p}_i(0) - \tilde{p}_e(0) = 0$, in the absence of the external electric field. With this aim let the initial perturbations are used (see also Figs. 4.1, 4.2):

```
v(0)=1,r(0)=1,s(0)=1/1838,u(0)=1,p(0)=1,q(0)=1,
D(v)(0)=0,D(r)(0)=0,D(s)(0)=0,D(u)(0)=0,D(p)(0)=0,D(q)(0)=0
```

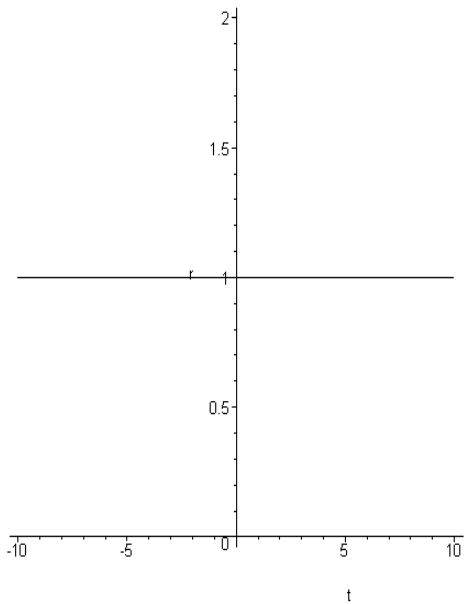

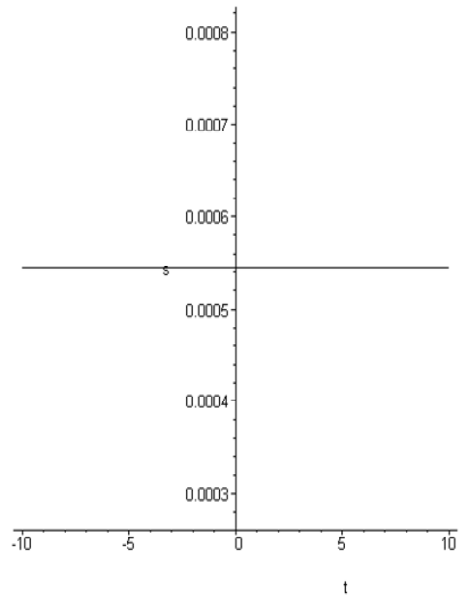

Fig. 4.1. r- density $\tilde{\rho}_i$.

Fig. 4.2. s- density $\tilde{\rho}_e$.

As expected Figs 4.1, 4.2 look like typical curves for free particles in the Schrödinger theory. But phonon - electron interaction does not lead to a chemical bonded system. It means that $\tilde{p}_i(0) - \tilde{p}_e(0) = 0$ and creation of the combined electron-phonon soliton can be realized in the superconducting system only in the external electric field posed by the lattice. In other words as result of polarization the "estafette" bond mechanism is appearing leading to wave solitons placed in bounded area of space.

As an example let us consider the situation when the soliton is catching by the external periodical non-potential longitudinal electric field

$$F_i^{(npot)} = \frac{eE}{m_i}\cos\left(kx - \omega t\right),\tag{4.1}$$

for which the phase velocity is $\omega / k = u_0$. In this case $\xi = x - Ct$, $\xi = x - u_0 t$ and (4.1) takes the form

$$F_i^{(npot)} = \frac{eE}{m_i}\cos\left[\frac{2\pi}{\lambda}x_0\tilde{\xi}\right],\tag{4.2}$$

But the value $v^{qu} = \hbar / m_e$ has the dimension $[\,cm^2/s\,]$ and can be titled as quantum viscosity, $v^{qu} = 1.1577 \; cm^2/s$. Introduce the quantum velocity $v^{qu} = \hbar / m_e$ in relation (4.2)

$$\rho_i F_i^{(npot)} = \rho_0 \tilde{\rho}_i \frac{eE}{m_i}\cos\left[\frac{2\pi}{\lambda}\frac{v^{qu}}{u_0}\tilde{\xi}\right].\tag{4.3}$$

Expression under cosines sign forms the similarity criteria which can be named as quantum Reynolds number, $\mathrm{Re}^{qu} = \dfrac{\lambda u_0}{v^{qu}}$. From (4.3) follows

$$\rho_i F_i^{npot} = \rho_0 \tilde{\rho}_i \frac{eE}{m_i}\cos\left[\frac{2\pi}{\mathrm{Re}^{qu}}\tilde{\xi}\right].\tag{4.4}$$

The effective force acting on the positive charges is written as

$$\rho_i F_i^{(pot)} + \rho_i F_i^{(npot)} = -\frac{u_0^2}{x_0}\frac{m_e}{m_i}\rho_0 \tilde{\rho}_i \left[\frac{\partial \tilde{\varphi}}{\partial \tilde{\xi}} - eE\frac{x_0}{m_e u_0^2}\cos\left(\frac{2\pi}{\mathrm{Re}^{qu}}\tilde{\xi}\right)\right].\tag{4.5}$$

Symbolize

$$\mathrm{E}^{qu} = eE\frac{x_0}{m_e u_0^2}\,.\tag{4.6}$$

and introduce $\mathrm{F}_0 = \dfrac{eE}{m_e}$ as the scale of the external force, acting on the mass unit of the positive charge, which the absolute value is $e$. Then

$$\mathrm{E^{qu}} = \frac{\mathrm{F_0} x_0}{u_0^2} \qquad (4.7)$$

is the similarity criteria reflecting the ratio the character work of the external force to the kinetic energy of the mass unit. We have

$$\rho_i F_i^{(pot)} + \rho_i F_i^{(npot)} = -\frac{u_0^2}{x_0} \frac{m_e}{m_i} \rho_0 \tilde{\rho}_i \left[ \frac{\partial \tilde{\varphi}}{\partial \tilde{\xi}} - \mathrm{E^{qu}} \cos\left( \frac{2\pi}{\mathrm{Re}^{qu}} \tilde{\xi} \right) \right]. \qquad (4.8)$$

Analogically for electrons

$$\rho_e F_e^{(pot)} + \rho_e F_e^{(npot)} = \frac{u_0^2}{x_0} \rho_0 \tilde{\rho}_e \left[ \frac{\partial \tilde{\varphi}}{\partial \tilde{\xi}} - \mathrm{E^{qu}} \cos\left( \frac{2\pi}{\mathrm{Re}^{qu}} \tilde{\xi} \right) \right]. \qquad (4.9)$$

Expressions (4.8), (4.9) should be introduced in the system of quantum hydrodynamic equations (3.21) - (3.26). The similarity criteria $\mathrm{E^{qu}}$ and $\mathrm{Re}^{qu}$ are parameters of calculations.

Let us show some results of calculations in external (and self-consistent) electric field for the following Cauchy conditions (written in Maple notations), (SYSTEM I):

$\mathrm{E^{qu}} = 1$, $\mathrm{Re}^{qu} = 1$, $\tilde{p}_e(0) = \tilde{p}_i(0)$,

<span style="color:red">v(0)=1,r(0)=1,s(0)=1/1838,u(0)=1,p(0)=1,q(0)=1,</span>

<span style="color:red">D(v)(0)=0,D(r)(0)=0,D(s)(0)=0,D(u)(0)=0,D(p)(0)=0,D(q)(0)=0.</span>

Figs. 4.3 – 4.6 reflect the result of calculation.

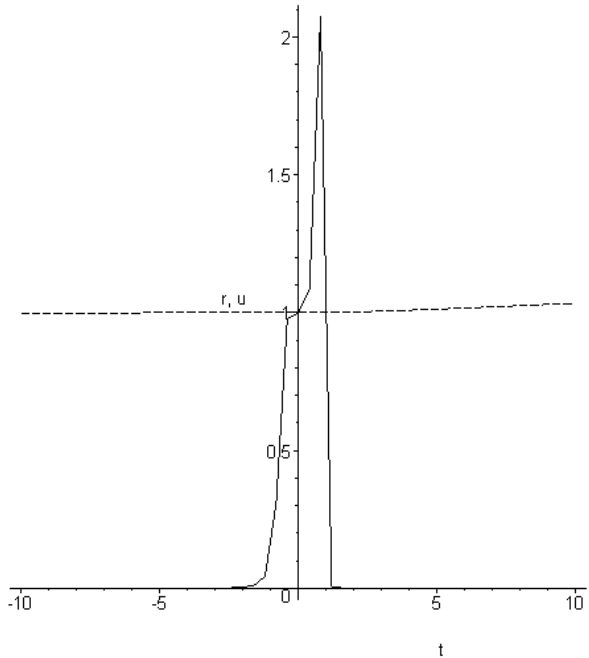
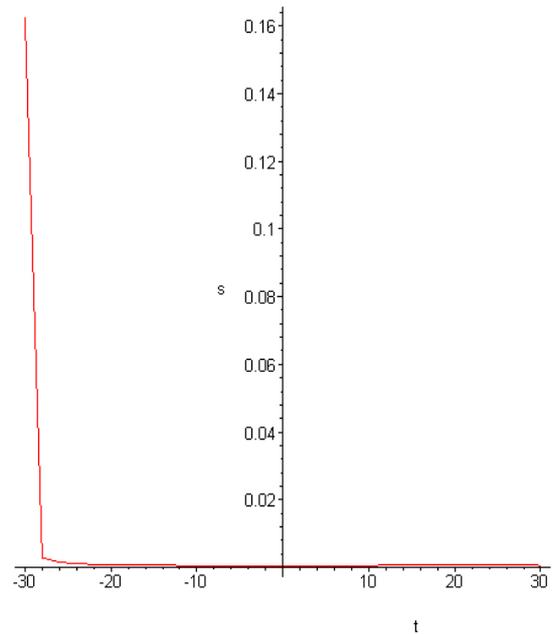

Fig. 4.3. r- density $\tilde{\rho}_i$, u- velocty $\tilde{u}$ .

Fig. 4.4. s- density $\tilde{\rho}_s$ .

]

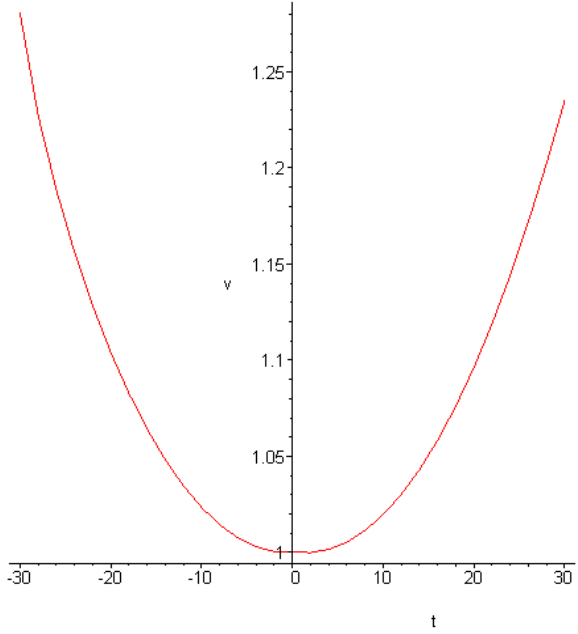

Fig. 4.5. v – self-consistent potential $\widetilde{\varphi}$ .

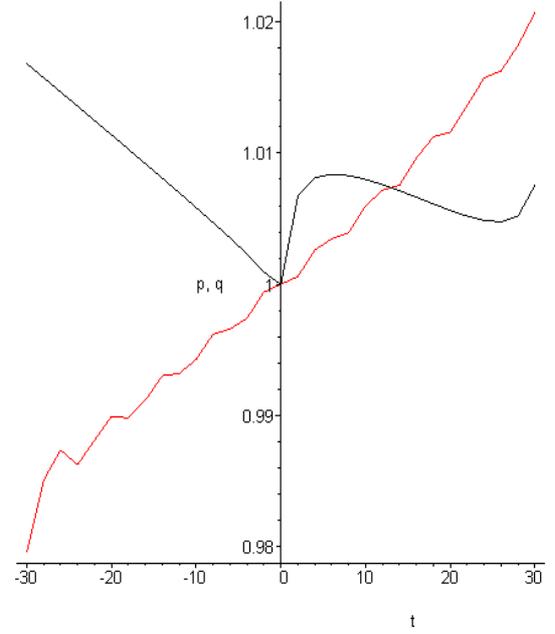

Fig. 4.6. p – pressure $\widetilde{p}_i$ ; q – pressure $\widetilde{p}_e$ ,

(wavy line)

The electric field of real lattices can have the complicated configuration. For analytical expression for forces created by ions of the crystal lattice, Fourier approximation can be used. Let us consider for example the influence of the quadratic harmonics (see (4.10), (4.11)) by the same other conditions of calculations (SYSTEM I). For the lattice ions

$$\rho_i F_i^{(pot)} + \rho_i F_i^{(npot)} = -\frac{u_0^2}{x_0}\frac{m_e}{m_i}\rho_0\widetilde{\rho}_i\left[\frac{\partial\widetilde{\varphi}}{\partial\widetilde{\xi}} - \mathrm{E}^{qu}\cos^2\left(\frac{2\pi}{\mathrm{Re}^{qu}}\widetilde{\xi}\right)\right]. \qquad (4.10)$$

Analogically for electrons

$$\rho_e F_e^{(pot)} + \rho_e F_e^{(npot)} = \frac{u_0^2}{x_0}\rho_0\widetilde{\rho}_e\left[\frac{\partial\widetilde{\varphi}}{\partial\widetilde{\xi}} - \mathrm{E}^{qu}\cos^2\left(\frac{2\pi}{\mathrm{Re}^{qu}}\widetilde{\xi}\right)\right]. \qquad (4.11)$$

The results of calculations on Figs. 4.7 – 4.10 are shown.

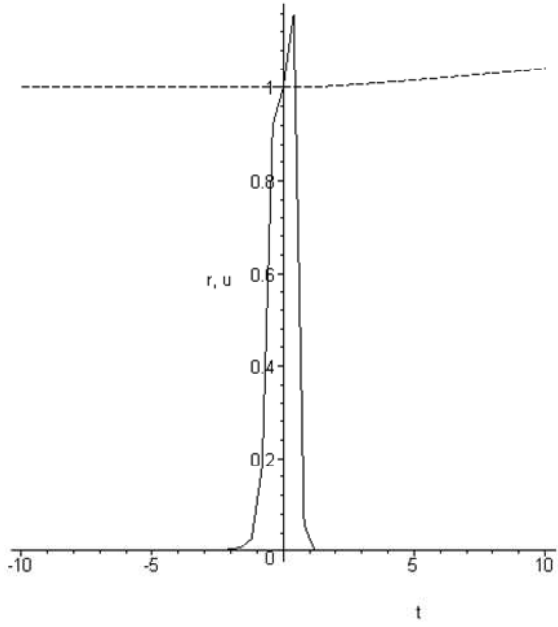

Fig. 4.7. r- density $\tilde{\rho}_i$, u- velocity $\tilde{u}$.

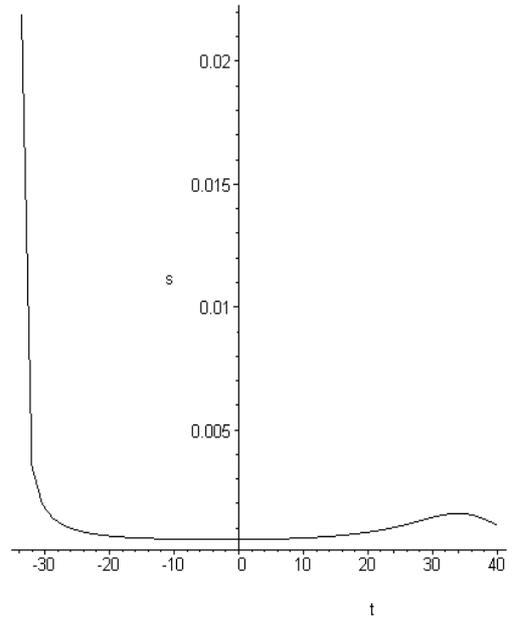

Fig. 4.8. s- density $\tilde{\rho}_s$.

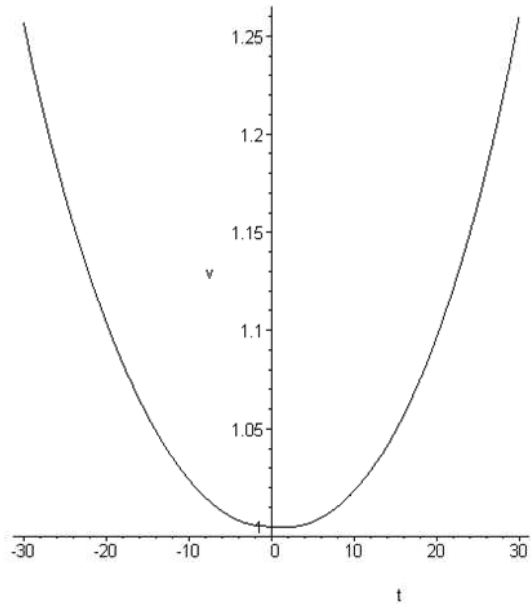

Fig. 4.9. v – self-consistent potential $\tilde{\varphi}$.

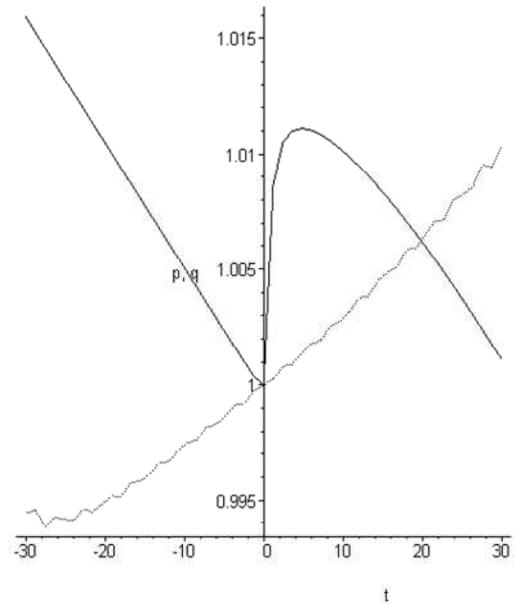

Fig. 4.10. p – pressure $\tilde{p}_i$, q – pressure $\tilde{p}_e$,

(wavy line)

The following results of calculations in external (and self-consistent) electric field correspond to Cauchy conditions (written in Maple notations), (SYSTEM 2):

$E^{qu}$ =1, $Re^{qu}$ =1, $\tilde{p}_e(0) = \tilde{p}_i(0)$,

```
v(0)=1,r(0)=1,s(0)=1/1838,u(0)=1,p(0)=2,q(0)=2,
D(v)(0)=0,D(r)(0)=0,D(s)(0)=0,D(u)(0)=0,D(p)(0)=0,D(q)(0)=0.
```

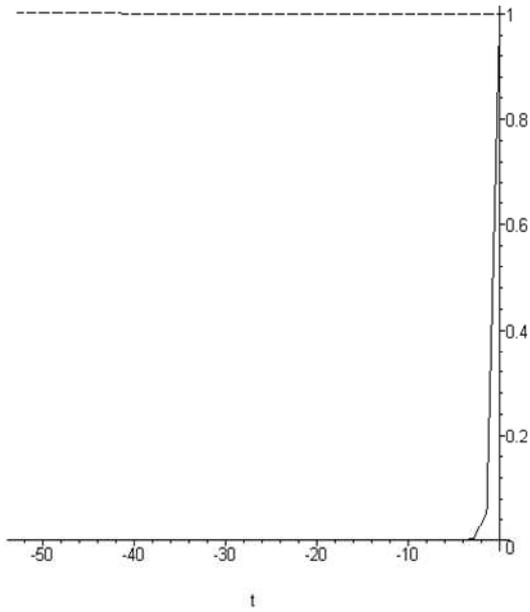

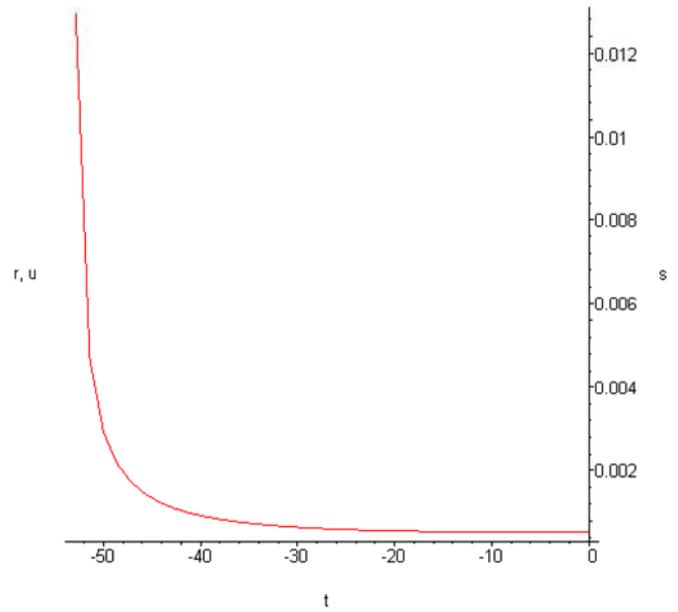

Fig. 4.11. r- density $\tilde{\rho}_i$, u- velocity $\tilde{u}$ .　　　　　Fig. 4.12. s- density $\tilde{\rho}_s$ .

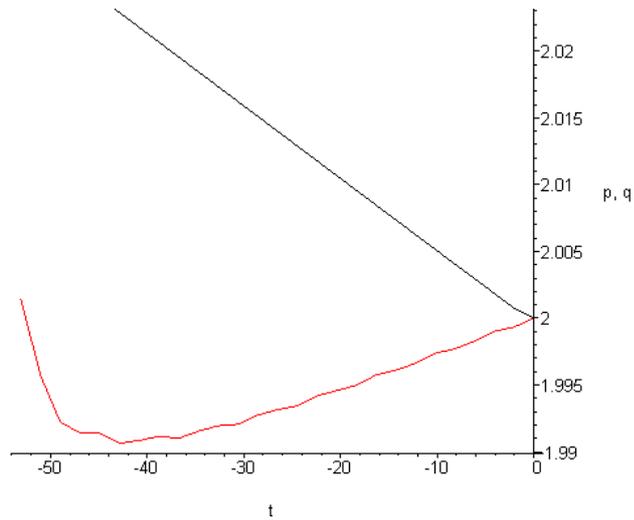

Fig. 4.13. p – pressure $\tilde{p}_i$ , q – pressure $\tilde{p}_e$ , (low line).

Some results of calculations in external (4.10), (4.11) (and self-consistent) electric field for the following Cauchy conditions (written in Maple notations), (SYSTEM 3):

$E^{qu}$ =2, $Re^{qu}$ =2, $\tilde{p}_e(0) = \tilde{p}_i(0)$,

```
v(0)=1,r(0)=1,s(0)=1/1838,u(0)=1,p(0)=1,q(0)=1,
D(v)(0)=0,D(r)(0)=0,D(s)(0)=0,D(u)(0)=0,D(p)(0)=0,D(q)(0)=0.
```

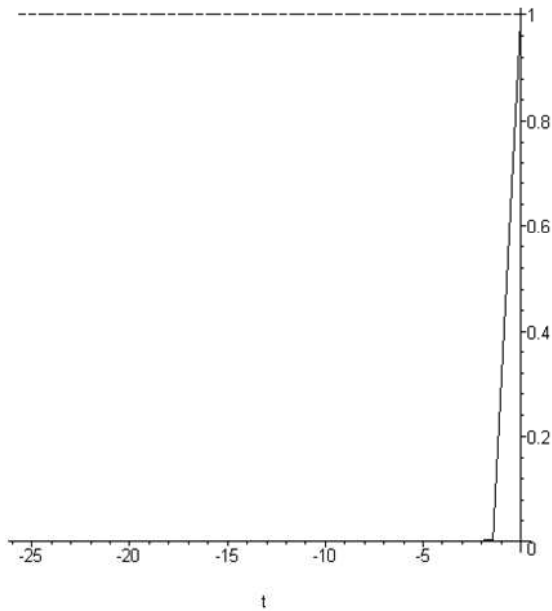

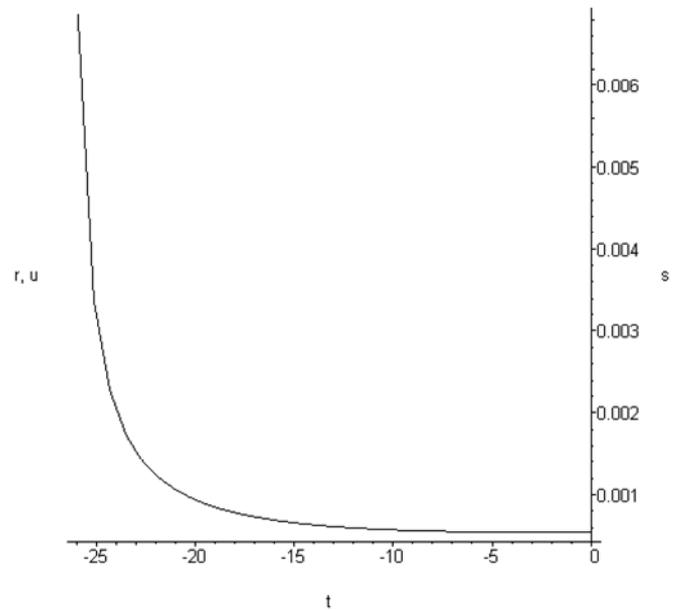

Fig. 4.14. r- density $\tilde{\rho}_i$, u- velocity $\tilde{u}$ .        Fig. 4.15. s- density $\tilde{\rho}_s$ .

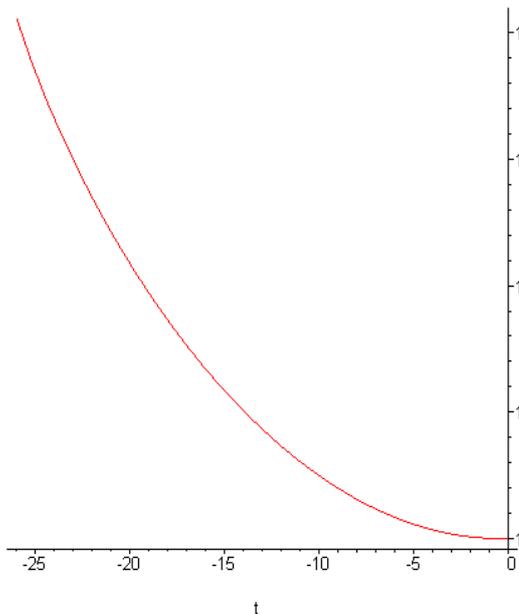

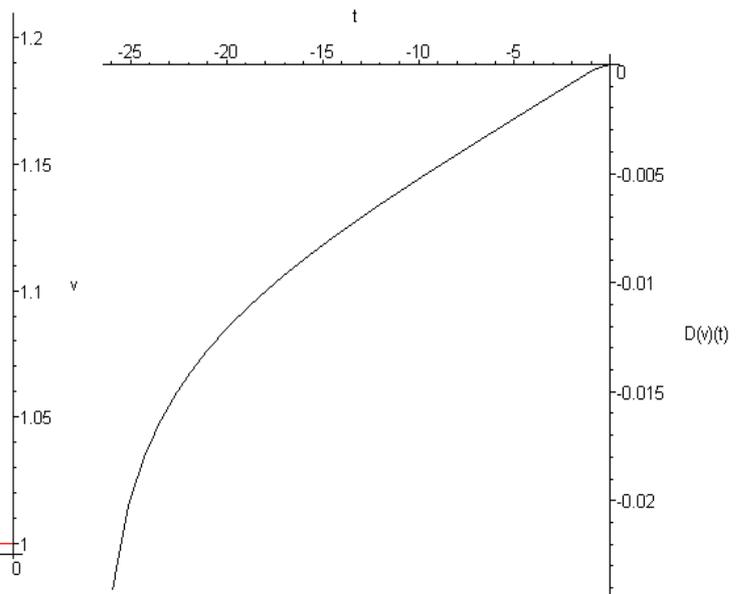

Fig. 4.16. v – self-consistent potential $\tilde{\varphi}$ .        Fig. 4.17. Derivative of the self-consistent potential $\tilde{\varphi}$ .

The previous figures display the typical quantum objects placed in *the bounded region* of 1D space, all parts of these objects are moving with the same velocity. Namely from calculations follow that in coordinate system moving with the phase velocity, indestructible soliton has the velocity equal to the phase velocity. Moreover the attempt to impose to soliton to move with another group velocity leads to the soliton destruction. As we see from calculations, in the superconductivity regime the "estafette" movement of the soliton system "lattice ion – electron" is realized without creation of a chemical bond. The dependences $\tilde{p}_e\left(\tilde{\xi}\right), \tilde{p}_i\left(\tilde{\xi}\right)$ define the conditions of the stability of the electron – ion pair. The destruction of the SC regime is the solitons destruction, depending on many conditions.

Research of superconductors is curried out very actively. But in spite of obvious success the following conclusion could be established:

1. Contemporary theories of superconductivity based on the Schrödinger equation, practically exhaust their arguments and have no possibility to explain effects of the high temperature superconductivity.

2. Contemporary theories of superconductivity (including BCS) based on the Schrödinger equation, can't propose the principles of search and creation of superconducting materials.

3. The necessity of creation of principal new non-local quantum theories of superconductivity is existing.

From position of the quantum hydrodynamics the problem of search and creation of superconductive materials come to the search of materials which lattices ensure the soliton movement without destruction. In my opinion the mentioned materials can be created artificially using the technology of the special introduction of quantum dots in matrices on the basement of proposed quantum hydrodynamics. It is known that technology of material creation with special quantum dots exists now in other applications.

## References.